\documentclass[aps,prd,twocolumn,superscriptaddress,showpacs,showkeys]{revtex4-1}
\usepackage{amsmath}
\usepackage{amsfonts}

\begin{document}

\title{General intrinsic theory of general large $N_{c}$ QCD, SU(3) QCD,
SU(2) hadron-dynamics and U(1) QED gauge field theories in general field
theory and progress towards solving the nucleon spin crisis}
\author{C. Huang}
\email{c.huang0@hotmail.com}
\affiliation{Department of Physics and Astronomy, Purdue University, 525 Northwestern Avenue, W.Lafayette, IN 47907-2036, USA}
 \author{Yong-Chang Huang}
 \email{ychuang@bjut.edu.cn}
\affiliation{Institute of Theoretical Physics, Beijing University of Technology,
 Beijing, 100124, China}
 \author{B. H. Zhou}
\affiliation{Department of Physics, Emory University, Atlanta, Georgia 30322-2430, USA}

\begin{abstract}
This paper gives general intrinsic theory of general large $N_{c}$ QCD,
SU(3) QCD, SU(2) hadron-dynamics and U(1) QED gauge field theories in
general field theory and progress towards solving the nucleon spin crisis,
i.e., presents general large $N_{c}$ QCD's inner structures, gauge invariant
angular momenta and new corresponding Coulomb theorem in quark-gluon field
interaction systems based on general field theory, and naturally deduces the
gauge invariant spin and orbital angular momentum operators of quark and
gauge fields with $SU(N_{c})$ gauge symmetry by Noether theorem in general
field theory. In the general large $N_{c}$ QCD, we discover not only the
general covariant transverse and parallel conditions ( namely, non-Abelian
divergence and curl ), but also that this general system has good intrinsic
symmetry characteristics. Specially, this paper's generally decomposing
gauge potential theory presents a new technique, it should play a votal role
in future physics research. Therefore, this paper breakthroughs the some
huge difficulties in the nucleon spin crisis and opens a door of researching
on lots of strong interacting systems with different symmetric properties,
which is popularly interesting, and keeps both the gauge invariance and
their angular momentum commutation relations so that their theories are
consistent. Especially, the achieved results here can be utilized to
calculate the general QCD strong interactions and to give the precise
predictions that can be exactly measured by current particle physics
experiments due to their gauge invariant properties etc.
\end{abstract}

\maketitle

1.Introduction

Yang and Mills, in 1954 for the first time, generalized the Abelian U(1)
gauge theory to non-Abelian SU(2) gauge theory to give a strong interaction
theory \cite{1}. This theory now has be greatly developed and is viewed as
the powerful and useful Yang-Mills gauge field theory. As a further
development of it, SU(2) hadron-dynamics gauge field theory in high energy
physics is the strong interaction theory of different states relevant to
different charged electricity states.

As is well known to people, up to know, the investigations on nuclear
theories have been awarded three Nobel physics prizes because of their huge
impacts for physics science. It is well-known that QCD is useful for
research on general strong interactions of quark-gluon systems, and large $%
N_{c}$ QCD is currently utilized to investigate quark-gluon interactions,
e.g., Ref.\cite{2} studied Nucleon-Antinucleon Annihilation in Large $N_{c}$
QCD.

The well-known nucleon spin crisis shows the key and urgent problem: how do
the motions of quarks and gluons in nucleon give the contribution to the
total spin of the nucleon? Various investigations show that the spin of the
quarks can only give one third of the nucleon spin \cite{3,4,5}, therefore,
the other part must be achieved from the other internal factors or motions:
the spin of the gluons and the orbital angular movements of the quarks and
gluons. In order to uncover how these motions give the contribution to the
total nucleon spin, the original work is generally to define the operators
of fully representing these movements. Various kinds of good endeavors have
been made for solving the crisis, e.g.,Refs.\cite{6,7,8,9,10,11}.

We, in this paper, generalize the relevant investigations in QED \cite{12}
and isospin gauge field theory \cite{13} in order to investigate the inner
structure of the interaction systems of the quark-gluon fields under the
intrinsic decomposition of color gauge fields. This kind of generalizations
is nontrivial, and people will find that there must be some required
additional conditions so that the new achieved theory is gauge invariant and
consistent. Furthermore, people will see that all the non-Abelian consistent
conditions can be impressively achieved by properly generalizing their
Abelian corresponding parts that had been exactly shown in Ref.\cite{12}.
Ref.\cite{14} had given two of the consistent conditions, while Ref.\cite{13}
further found that the other four consistent conditions show the properties
of different components of the gauge field strength in SU(2) isospin gauge
field theory. Because, in strong interaction systems with isospin symmetry,
there still is the serious problem that gauge invariant angular momenta are
still missing \cite{15,16}, Ref.\cite{13} solved the serious problem. This
paper wants not only to generalize the solutions to the key problems,
analogous to nucleon spin crisis, in QED \cite{12} and SU(2) isospin gauge
theory \cite{13}, respectively, ( namely, based on the solid bases of Refs.%
\cite{12,13} at al ) in order to make essential steps towards solving
nucleon spin crisis, but also to give their consistent unification
description theory.

The arrangement of this paper: Sect. 2 is $SU(N_{c})$ gauge field inner
structures and new Jacobi identity; Sect. 3 is general decompositions of $%
SU(N_{c})$ gauge fields; Sect. 4 shows gauge invariant angular momenta of $%
SU(N_{c})$ quark-gluon field interaction systems and gives a new viewpoint
on the resolution of the nucleon spin crisis; Sect. 5 is discussion; Sect. 6
gives summary and conclusion.

2. $SU(N_{c})$ gauge field inner structures and new Jacobi identity

For $SU(N_{c})$ gauge field generators $T^{a},a=1,2,...,N_{c}^{2}-1;N_{c}$
may be enough large, we have
\begin{equation*}
T^{a}T^{b}=\frac{1}{2}(T^{a}T^{b}+T^{b}T^{a})+\frac{1}{2}%
(T^{a}T^{b}-T^{b}T^{a})
\end{equation*}%
\begin{equation}
=\frac{1}{2N_{c}}\delta ^{ab}+\frac{1}{2}%
(d^{abe}+if^{abe})T^{e},N_{1}=N,N_{2}=M,  \tag{2.1}
\end{equation}%
because $\{T^{a},T^{b}\}=\frac{1}{2N_{c}}\delta ^{ab}+\frac{1}{2}%
d^{abe}T^{e},[T^{a},T^{b}]=if^{abe}T^{e}$. Using Eq.(2.1), we have\quad
\quad \quad \quad \quad
\begin{equation}
\frac{1}{2N_{c}}\delta ^{ab}T^{c}T^{d}=[T^{a}T^{b}-\frac{1}{2}%
(d^{abe}+if^{abe})T^{e}]T^{c}T^{d}.  \tag{2.2}
\end{equation}%
Similar to deduction of Eq.(2.2), we get\quad \quad \quad \quad
\begin{equation}
\frac{1}{2N_{c}}\delta ^{ad}T^{b}T^{c}=[T^{a}T^{d}-\frac{1}{2}%
(d^{ade}+if^{ade})T^{e}]T^{b}T^{c},  \tag{2.3}
\end{equation}%
Using Eqs.(2.2) and (2.3), we can have\quad
\begin{equation*}
\frac{1}{4N_{c}}(\delta ^{ab}\delta ^{c}{}^{d}-\delta ^{ad}\delta
^{b}{}^{c})=tr(T^{a}[T^{b}T^{c},T^{d}])-\frac{1}{2}(d^{abe}+if^{abe})\frac{1%
}{4}(
\end{equation*}%
\begin{equation}
d^{ecd}+if^{ecd})+\frac{1}{2}(d^{ade}+if^{ade})\frac{1}{4}(d^{ecd}+if^{ecd})
\tag{2.4}
\end{equation}%
where we have used $tr(T^{a}T^{b})=\frac{1}{2}\delta ^{ab}$ and $%
tr(T^{a}T^{b}T^{c})=\frac{1}{4}(d^{abc}+if^{abc})$ \cite{16}.

Using $[T^{b}T^{c},T^{d}]=T^{b}[T^{c},T^{d}]+[T^{b},T^{d}]T^{c}$ and further
simplifying Eq.(2.4), we deduce

\begin{equation*}
\frac{2}{N_{i}}(\delta ^{ab}\delta ^{c}{}^{d}-\delta ^{ad}\delta
^{b}{}^{c})+d^{abe}d^{c}{}^{de}=f^{ace}f^{bde}+d^{ade}d^{b}{}^{ce}+
\end{equation*}

\begin{equation}
i(d^{abe}f^{cde}+2d^{ace}f^{bde}-f^{abe}d^{ecd}+f^{ade}d^{ebc}+d^{ade}f^{ebc})
\tag{2.5}
\end{equation}

Using Eq.(2.5), we achieve both a general contraction expression of
structure constants of $SU(N_{c})$ gauge group

\begin{equation}
f^{ace}f^{bde}=\frac{2}{N_{c}}(\delta ^{ab}\delta ^{c}{}^{d}-\delta
^{ad}\delta ^{b}{}^{c})+d^{abe}d^{c}{}^{de}-d^{ade}d^{b}{}^{ce}  \tag{2.6}
\end{equation}%
and an equation

\begin{equation}
d^{abe}f^{cde}+2d^{ace}f^{bde}-f^{abe}d^{ecd}+f^{ade}d^{ebc}+d^{ade}f^{ebc}=0
\tag{2.7}
\end{equation}

Using Jacobi identity $d^{abe}f^{cde}+d^{ace}f^{bde}+f^{ade}d^{ebc}=0$ \cite%
{16} to simplify Eq.(2.7), we deduce a new Jacobi identity of structure
constants

\begin{equation}
d^{ace}f^{bde}-f^{abe}d^{ecd}+d^{ade}f^{ebc}=0,  \tag{2.8}
\end{equation}%
because Eq.(2.8) can directly be deduced by

\begin{equation}
\{T^{a},[T^{b},T^{c}]\}-[T^{b},\{T^{c},T^{a}\}]-\{T^{c},[T^{a},T^{b}]\}=0
\tag{2.9}
\end{equation}%
Eq.(2.9) is just a new Jacobi identity, which can be proved by directly
expanding Eq.(2.9). Eq.(2.8) gives a new relation between symmetric and
antisymmetric structure constants.

The other more investigations on inner structures of $SU{}(N_{c})$ gauge
fields will be naturally given in following sections.

3. General decompositions of $SU(N_{c})$ gauge fields

As a general generalization of the QED and the isospin cases, we investigate
general decompositions of $SU(N_{c})$ gauge fields by the general method of
field theory. The projection operators are defined as \cite{12,13}
\begin{equation}
L_{k}^{j}=\partial \strut ^{j}\frac{1}{\Delta }\partial
_{k},T_{k}^{j}=\delta _{k}^{j}-\strut L_{k}^{j},(\Delta =\partial
_{k}\partial ^{k})  \tag{3.1}
\end{equation}%
and the gauge potential $A^{ia}(i=1,2,3;a=1,2,3,...,N_{c}^{2}-1)$ are
intrinsically decomposed as
\begin{equation}
A_{\bot }^{ja}=T_{k}^{j}A^{ka},A_{\Vert }^{ja}=L_{k}^{j}A^{ka}.  \tag{3.2}
\end{equation}

The general action for the $SU(N_{c})$ quark-gluon field interaction system
is \cite{16}
\begin{equation}
S=\int d^{4}x[-\frac{1}{4}F^{a\mu \nu }F_{a\mu \nu }+\overline{\psi }\left(
i\gamma ^{\mu }D_{\mu }-m\right) \psi ],  \tag{3.3}
\end{equation}%
where $D_{\mu }=\partial _{\mu }-igA_{\mu }$ are the covariant derivative
with gauge fields $A_{\mu }=A_{\mu }^{a}T^{a}$ and $T^{a}$ the generator of
the $SU(N_{c})$ group, and $\psi $ stands for the quark field. When $D_{\mu
} $ acts on any vector in Lie algebra space, we have $D_{\mu }B=\partial
_{\mu }B^{a}T^{a}-ig[A_{\mu },B^{a}T^{a}]=$ $(\partial _{\mu
}B^{a}+gf^{abc}A_{\mu }^{b}B^{c})T^{a}.$ In $SU(N_{c})$ gauge field theory,
the gauge field strength is
\begin{equation}
F^{a\mu \nu }=\partial ^{\mu }A^{a\nu }-\partial ^{\nu }A^{a\mu
}+gf^{abc}A^{b\mu }A^{c\nu }.  \tag{3.4}
\end{equation}

We can decompose the spacial part of the gauge potential $A^{ak}$ into two
components
\begin{equation}
A^{ak}=A_{\bot }^{ak}+A_{\Vert }^{ak}.  \tag{3.5}
\end{equation}%
Analogous to QED \cite{12}, under a $SU(N_{c})$ gauge transformation $U$, $%
A_{\bot }^{k}=A_{\bot }^{ak}T^{a}$ and $A_{\Vert }^{k}=A_{\Vert }^{ak}T^{a}$
transform as \cite{13}
\begin{equation}
A_{\bot }^{^{\prime }k}=UA_{\bot }^{k}U^{\dagger },  \tag{3.6}
\end{equation}%
\begin{equation}
A_{\Vert }^{^{\prime }k}=UA_{\Vert }^{k}U^{\dagger }-\frac{i}{g}U\partial
^{k}U^{\dagger },  \tag{3.7}
\end{equation}%
respectively, which make
\begin{equation}
A^{^{\prime }k}=UA^{k}U^{\dagger }-\frac{i}{g}U\partial ^{k}U^{\dagger }.
\tag{3.8}
\end{equation}%
The expressions from Eq.(3.5) to Eq.(3.8) simplify the relevant expressions
in Ref.\cite{13}.

Using Eqs.(3.2) and (3.5), the general action (3.3) can be rewritten as

\begin{equation*}
S=\int d^{4}x[-\frac{1}{2}(\partial ^{k}A^{a0}\partial
_{k}A_{0}^{a}+\partial ^{0}A_{\Vert }^{ak}\partial _{0}A_{\Vert
k}^{a}+\partial ^{0}A_{\bot }^{ak}\partial _{0}A_{\bot k}^{a})
\end{equation*}

\begin{eqnarray*}
&&+\partial ^{k}A^{a0}\partial _{0}A_{\Vert k}^{a}+\partial
^{k}A^{a0}\partial _{0}A_{\bot k}^{a}+\partial ^{0}A_{\Vert }^{ak}\partial
_{0}A_{\bot k}^{a}-\partial ^{k}A^{a0}( \\
&&A_{\Vert k}^{b}A_{0}^{c}+A_{\bot k}^{b}A_{0}^{c})gf^{abc}+\partial
^{0}A_{\Vert }^{ak}gf^{abc}(A_{\Vert k}^{b}A_{0}^{c}+A_{\bot
k}^{b}A_{0}^{c})+ \\
&&\partial ^{0}A_{\bot }^{ak}gf^{abc}(A_{\Vert k}^{b}A_{0}^{c}+A_{\bot
k}^{b}A_{0}^{c})-\frac{1}{2}gf^{abc}(A_{\Vert }^{bk}A^{c0}+A_{\bot
}^{bk}A^{c0})(
\end{eqnarray*}

\begin{equation}
A_{\Vert k}^{b^{^{\prime }}}A_{0}^{c^{^{\prime }}}+A_{\bot k}^{b^{^{\prime
}}}A_{0}^{c^{^{\prime }}})gf^{ab^{^{\prime }}c^{^{\prime }}}-\frac{%
F^{ajk}F_{jk}^{a}}{4}\ +\overline{\psi }\left( i\gamma ^{\mu }D_{\mu
}-m\right) \psi ]  \tag{3.9}
\end{equation}

In order to cancel the non-independent terms in Eq.(3.9), we need to use the
basic consistent conditions in $SU(N_{c})$ gauge theory to simplify Eq.(3.9).

Because the nonlinearity of the non-Abelian $SU(N_{c})$ gauge field system,
the non-Abelian system is much more complex than the Abelian one. We now
generalize the two well known Abelian gauge potential (or field) consistence
conditions in QED \cite{12}
\begin{equation}
\nabla \cdot \overset{\rightharpoonup }{A}_{\bot }=0,  \tag{3.10}
\end{equation}

\begin{equation}
\nabla \times \overset{\rightharpoonup }{A}_{\Vert }=0,  \tag{3.11}
\end{equation}%
respectively to non-Abelian $SU(N_{c})$ corresponding expressions
\begin{equation}
\partial _{k}A_{\bot }^{ak}+gf^{abc}A_{\Vert k}^{b}A_{\bot }^{ck}=0.
\tag{3.12}
\end{equation}%
\begin{equation}
(D_{\Vert }^{j}D_{\Vert }^{k}-D_{\Vert }^{k}D_{\Vert
}^{j})^{a}=0,i.e.\partial ^{j}A_{\Vert }^{ak}-\partial ^{k}A_{\Vert
}^{aj}+gf^{abc}A_{\Vert }^{bj}A_{\Vert }^{ck}=0  \tag{3.13}
\end{equation}%
One can easily prove that Eqs.(3.12) and (3.13) are $SU(N_{c})$ gauge
covariant, and they are similar to the isospin relevant expressions in Ref.%
\cite{13} except the structure constants and superscripts' taking values of $%
SU(N_{c})$ gauge group and being related to gauge fields \cite{16}.

Using Eq.(3.12) we can have%
\begin{equation*}
\int dx^{3}\partial ^{k}A^{a0}\partial _{0}A_{\bot k}^{a}-\int
dx^{3}gf^{abc}\partial _{0}A_{\Vert k}^{a}A_{\bot }^{bk}A^{c0}
\end{equation*}

\begin{equation}
+\int dx^{3}gf^{abc}\partial _{0}A_{\bot }^{ak}A_{\Vert k}^{b}A^{c0}=0
\tag{3.14}
\end{equation}

where the divergence term has been neglected.

\ \ \ We further utilize the consistent conditions (whose physics meanings
will be explained latter)

\begin{equation}
\int d^{3}x\partial ^{0}A_{\Vert }^{ak}(-\partial _{0}A_{\bot
k}^{a}+gf^{abc}A_{\bot k}^{b}A^{c0})=0,  \tag{3.15}
\end{equation}

\begin{eqnarray*}
&&gf^{abc}(\partial _{0}A_{\Vert }^{ak}A_{\bot k}^{b}A^{c0}-\partial
^{k}A^{a0}A_{\bot k}^{b}A_{0}^{c}) \\
&&-\frac{g^{2}2}{N_{c}}(A_{\Vert }^{bk}A^{c0}A_{\bot
k}^{b}A_{0}^{c}-A_{\Vert }^{bk}A^{c0}A_{\bot k}^{c}A_{0}^{b})
\end{eqnarray*}%
\begin{equation}
-g^{2}\left( d^{abb^{\prime }}d^{acc^{\prime }}-d^{abc^{\prime
}}d^{acb^{\prime }}\right) A_{\Vert }^{bk}A^{c0}A_{\bot k}^{b^{^{\prime
}}}A_{0}^{c^{^{\prime }}}=0,  \tag{3.16}
\end{equation}%
and substitute the contraction expression (2.6) of structure constants of $%
SU(N_{c})$ gauge group into Eq.(3.9), we achieve

\begin{eqnarray*}
&&S=\int d^{4}x\{-\frac{1}{2}(\partial ^{k}A^{a0}\partial
_{k}A_{0}^{a}+\partial ^{0}A_{\Vert }^{ak}\partial _{0}A_{\Vert
k}^{a}+\partial ^{0}A_{\bot }^{ak}\partial _{0}A_{\bot k}^{a}) \\
&&+\partial ^{k}A^{a0}\partial _{0}A_{\Vert k}^{a}-\partial
^{k}A^{a0}(gf^{abc}A_{\Vert k}^{b}A_{0}^{c}+gf^{abc}A_{\bot k}^{b}A_{0}^{c})+
\\
&&\partial ^{0}A_{\Vert }^{ak}gf^{abc}A_{\Vert k}^{b}A_{0}^{c}+\partial
^{0}A_{\bot }^{ak}gf^{abc}A_{\bot k}^{b}A_{0}^{c}-\frac{g^{2}}{N_{c}}%
(A_{\Vert }^{bk}A_{\Vert k}^{b}A^{c0}A_{0}^{c} \\
&&-A_{\Vert }^{bk}A_{0}^{b}A_{\Vert k}^{c}A^{c0}+A_{\bot }^{bk}A_{\bot
k}^{b}A^{c0}A_{0}^{c}-A_{\bot }^{bk}A_{0}^{b}A_{\bot k}^{c}A^{c0})- \\
&&\frac{1}{2}g^{2}\left( d^{abb^{\prime }}d^{acc^{\prime }}-d^{abc^{\prime
}}d^{acb^{\prime }}\right) (A_{\Vert }^{bk}A^{c0}A_{\Vert k}^{b^{^{\prime
}}}A_{0}^{c^{^{\prime }}}+
\end{eqnarray*}

\begin{equation}
A_{\bot }^{bk}A^{c0}A_{\bot k}^{b^{^{\prime }}}A_{0}^{c^{^{\prime }}})-\frac{%
1}{4}F^{ajk}F_{jk}^{a}\ +\overline{\psi }\left( i\gamma ^{\mu }D_{\mu
}-m\right) \psi \}  \tag{3.17}
\end{equation}

One can see that there is no the cross terms between parallel and vertical
gauge potentials in Eq.(3.17), which just shows that the general system has
very good symmetry properties, i.e., we discover that this general system
has very good intrinsic symmetry characteristics.

Therefore, we can calculate the corresponding canonical momenta conjugate to
the two components of the gauge potential, respectively,

\begin{equation}
\pi _{\Vert }^{ak}=\frac{\delta L}{\delta (\partial _{0}A_{\Vert k}^{a})}%
=-\partial ^{0}A_{\Vert }^{ak}+\partial ^{k}A^{a0}+gf^{abc}A_{\Vert
}^{bk}A^{c0}  \tag{3.18}
\end{equation}

\begin{equation}
\pi _{\bot }^{ak}=\frac{\delta L}{\delta (\partial _{0}A_{\bot k}^{a})}%
=-\partial ^{0}A_{\bot }^{ak}+gf^{abc}A_{\bot }^{bk}A^{c0}  \tag{3.19}
\end{equation}%
Eqs.(3.18) and (3.19) are the direct generalization of the isospin relevant
expressions in Ref.\cite{13} with the structure constants and superscripts'
taking values of $SU(N_{c})$ gauge group and are related to gauge fields
\cite{14}. It is apparent that Eqs.(3.18) and (3.19) are both gauge
covariant and their summation is exactly the conventional $\pi ^{ak}$, which
show that the above investigations are consistent.

Consequently, we need further to generalize, in QED respectively, the
well-known Abelian gauge field strength consistent conditions \cite{14}

\begin{equation}
\nabla \cdot \overset{\rightharpoonup }{E_{\bot }}=0,  \tag{3.20}
\end{equation}

\begin{equation}
\nabla \times \overset{\rightharpoonup }{E_{\Vert }}=0,  \tag{3.21}
\end{equation}%
\begin{equation}
\nabla \cdot \overset{\rightharpoonup }{E_{\Vert }}=\rho _{e},  \tag{3.22}
\end{equation}%
to non-Abelian $SU(N_{c})$ corresponding expressions
\begin{equation}
\partial _{k}\pi _{\bot }^{ak}+gf^{abc}A_{\Vert k}^{b}\pi _{\bot }^{ck}=0.
\tag{3.23}
\end{equation}%
\begin{equation}
D^{j}\pi _{\Vert }^{ak}-D^{k}\pi _{\Vert }^{aj}=0,  \tag{3.24}
\end{equation}%
\begin{equation}
\partial _{k}\pi _{\Vert }^{ak}+gf^{abc}A_{\Vert k}^{b}\pi _{\Vert
}^{ck}=\rho ^{a}=g\psi ^{\dagger }T^{a}\psi ,  \tag{3.25}
\end{equation}%
where $\rho ^{a}$ is the non-Abelian charge density.

4. Gauge invariant angular momenta of $SU(N_{c})$ quark-gluon field
interaction systems and a new viewpoint on the resolution of the nucleon
spin crisis

Using Noether theorem, the action (3.17) results in the angular momentum of
the $SU(N_{c})$ quark-gluon field interaction system as follows
\begin{equation*}
\overset{\rightharpoonup }{J_{1}}=\int d^{3}x\psi ^{\dagger }\frac{1}{2}%
\overset{\rightharpoonup }{\Sigma }\psi +\int d^{3}x\psi ^{\dagger }\overset{%
\rightharpoonup }{x}\times \frac{1}{i}\nabla \psi +\int d^{3}x\overset{%
\rightharpoonup }{\pi _{\bot }^{a}}\times \overset{\rightharpoonup }{A_{\bot
}^{a}}+
\end{equation*}

\begin{equation}
\int d^{3}x\overset{\rightharpoonup }{\pi _{\Vert }^{a}}\times \overset{%
\rightharpoonup }{A_{\Vert }^{a}}+\int d^{3}x\pi _{\bot k}^{a}\overset{%
\rightharpoonup }{x}\times \nabla A_{\bot }^{ak}+\int d^{3}x\pi _{\Vert
k}^{a}\overset{\rightharpoonup }{x}\times \nabla A_{\Vert }^{ak}.  \tag{4.1}
\end{equation}

Actually, Eq.(4.1) is a straightforward generalization that in Ref.\cite{13}.

In order to simplify Eq.(4.1), using Eq.(3.13) we have

\begin{equation*}
\nabla \cdot \lbrack \overset{\rightharpoonup }{\pi _{\Vert }^{a}}(\overset{%
\rightharpoonup }{A_{\Vert }^{a}}\times \overset{\rightharpoonup }{x})]=-%
\overset{\rightharpoonup }{\pi _{\Vert }^{a}}\times \overset{\rightharpoonup
}{A_{\Vert }^{a}}-\pi _{\Vert i}^{a}\overset{\rightharpoonup }{x}\times
\nabla A_{\Vert }^{ai}
\end{equation*}

\begin{equation}
-(\partial _{k}\pi _{\Vert }^{ak}+gf^{abc}A_{\Vert k}^{b}\pi _{\Vert }^{ck})(%
\overset{\rightharpoonup }{x}\times \overset{\rightharpoonup }{A_{\Vert }^{a}%
}).  \tag{4.2}
\end{equation}

Utilizing the divergence consistency condition for parallel components
(3.25) of gauge field strength and the identity (4.2), we can simplify
Eq.(4.1) as
\begin{eqnarray*}
\overset{\rightharpoonup }{J_{2}} &=&\int d^{3}x\psi ^{\dagger }\frac{1}{2}%
\overset{\rightharpoonup }{\Sigma }\psi +\int d^{3}x\psi ^{\dagger }\overset{%
\rightharpoonup }{x}\times \frac{1}{i}\overset{\rightharpoonup }{D_{\Vert }}%
\psi \\
&&+\int d^{3}x\overset{\rightharpoonup }{\pi _{\bot }^{a}}\times \overset{%
\rightharpoonup }{A_{\bot }^{a}}+\int d^{3}x\pi _{\bot k}^{a}\overset{%
\rightharpoonup }{x}\times \nabla A_{\bot }^{ak}
\end{eqnarray*}%
\begin{equation}
=\overset{\rightharpoonup }{S^{q}}+\overset{\rightharpoonup }{L_{2}^{q}}+%
\overset{\rightharpoonup }{S_{2}^{g}}+\overset{\rightharpoonup }{L_{2}^{g}},
\tag{4.3}
\end{equation}%
in which, because of Eq.(3.7), $\overset{\rightharpoonup }{D_{\Vert }}%
=\nabla -ig\overset{\rightharpoonup }{A_{\Vert }}$ is the new covariant
derivative. $\overset{\rightharpoonup }{S^{q}}$, $\overset{\rightharpoonup }{%
L_{2}^{q}}$, $\overset{\rightharpoonup }{S_{2}^{g}}$ and $\overset{%
\rightharpoonup }{L_{2}^{g}}$ are the spin and orbital angular momenta of
the quark field and gauge field, respectively. Eq.(4.3) is just a direct
generalization of the relevant expression of SU(2) isospin symmetry theory
in Ref.\cite{13} because of their similar mathematical structure.

Due to the properties of $A_{\Vert }^{ak}$ shown in Eqs.(3.7) and (3.13), $%
\overset{\rightharpoonup }{L_{2}^{q}}$ satisfies the commutation law $%
\overset{\rightharpoonup }{L_{2}^{q}}\times \overset{\rightharpoonup }{%
L_{2}^{q}}=i\overset{\rightharpoonup }{L_{2}^{q}}$ and is gauge invariant
\cite{6}. The gauge invariance of $\overset{\rightharpoonup }{S^{q}}$ and $%
\overset{\rightharpoonup }{S_{2}^{g}}$ is obvious; furthermore, $A_{\bot
}^{ak}$ is gauge covariant in Eq.(4.3), which means that $\overset{%
\rightharpoonup }{L_{2}^{g}}$ is gauge invariant.

Under the general Lorentz transformation of the fundamental fields,
Noether's theorem requires the invariance of the Lagrangian and Hamiltonian
densities of the general system \cite{17, 18}, e.g., this system is
invariant under the Lorentz transformations of $A^{\mu }$'s all components,
although Lorentz transformation of $A_{\perp }^{\mu }$ is complex, the
relevant complex terms of the transformations of all their components can be
canceled each other in the whole system according to the symmetric
invariance property of this system, which is the general rule, see Refs.\cite%
{17,18}. Furthermore, the frame-dependence issue has important physics
meaning, Refs.\cite{9,10,11} have given the very good investigations on the
issue etc.

Next, we will derive some useful relations from the consistency conditions
for field strength. Making gauge transformation of Eq.(4.3)'s the last term,
we have

\begin{equation*}
tr(\pi _{\bot k}^{\prime }\overset{\rightharpoonup }{x}\times \nabla A_{\bot
}^{\prime k})=tr(\pi _{\bot k}\overset{\rightharpoonup }{x}\times \nabla
A_{\bot }^{k})
\end{equation*}

\begin{equation}
+tr(x\times U^{+}\nabla U[A_{\bot k},\pi _{\bot }^{k}])=0,  \tag{4.4}
\end{equation}%
thus gauge invariance of the last term of Eq.(4.3) demands

\begin{equation}
f^{abc}A_{\bot k}^{b}\pi _{\bot }^{ck}=0,  \tag{4.5}
\end{equation}%
consequently we naturally deduce a new consistent condition \cite{6}.

Then, with the well known Coulomb law in QCD \cite{16}
\begin{equation}
\partial _{k}\pi ^{ak}+gf^{abc}A_{k}^{b}\pi ^{ck}=\rho ^{a},  \tag{4.6}
\end{equation}%
and in terms of Eqs.(3.23), (3.25) and (4.6), we get
\begin{equation}
gf^{abc}A_{\bot k}^{b}\pi ^{ck}=0.  \tag{4.7}
\end{equation}%
Using the consistent condition (4.5) of gauge invariance of the last term of
Eq.(4.3), we can simplify Eq.(4.7) as

\begin{equation}
f^{abc}A_{\bot k}^{b}\pi _{\Vert }^{ck}=0,  \tag{4.8}
\end{equation}%
which can be used to cancel some extra freedoms of this system.

Furthermore, in order to illustrate the meanings of the two consistent
conditions (3.15) and (3.16) in simplifying the original action,
substituting Eq.(3.19) into Eq.(3.15) and using Eq.(3.18), we show that
Eq.(3.15) is just the orthogonal relation between the two components of the
canonical momentum

\begin{equation}
\int d^{3}x\pi _{\Vert }^{ak}\pi _{\bot k}^{a}=0.  \tag{4.9}
\end{equation}%
which is just a generalization of Abelian orthogonal relation $\int
d^{3}xE_{\bot }^{k}E_{\Vert k}^{{}}=0$ (in the sense of the whole space) to
a non-Abelian case.

Then, substituting Eq.(3.19) into Eq.(3.23) and using Eqs.(2.6) and (3.12),
we can prove that

\begin{equation*}
gf^{abc}\partial ^{0}A_{\Vert k}^{b}A_{\bot }^{ck}+gf^{abc}A_{\bot
}^{bk}\partial _{k}A^{c0}-g^{2}\frac{2}{N_{c}}(A_{\Vert k}^{b}A_{\bot
}^{bk}A^{a0}+
\end{equation*}%
\begin{equation}
A_{\Vert k}^{a}A_{\bot }^{ck}A^{c0})+g^{2}(d^{abc}d^{b^{\prime }a^{\prime
}c}{}-d^{ab^{\prime }c}d^{b}{}^{a^{\prime }c})A_{\Vert k}^{b}A_{\bot
}^{a^{\prime }k}A^{b^{\prime }0}=0  \tag{4.10}
\end{equation}

Multiplying its right side by $A_{0}^{a}$ and rearranging the $SU(N_{c})$
group indices, Eq.(4.10) results in Eq.(3.16).

Consequently, all the assumptions being made here are consistent, namely,
Eqs.(3.15), (3.16) and (4.5) can be given out from Eqs.(3.12), (3.23),
(3.25) and (4.9), which are just directly generalized from the Abelian
conditions to non-Abelian cases in order to keep their gauge covariant way
and their consistent properties.

Furthermore, we will show the relation between our result and the previous
ones. To do this, we need to give another expression of angular momenta $%
J_{2}$. Using Eqs.(3.12), (3.13) and (4.8), we achieve the surface term

\begin{equation*}
\nabla \cdot \lbrack \overset{\rightharpoonup }{A_{\bot }^{a}}(\overset{%
\rightharpoonup }{\pi _{\Vert }^{a}}\times \overset{\rightharpoonup }{x})]=%
\overset{\rightharpoonup }{\pi _{\Vert }^{a}}\times \overset{\rightharpoonup
}{A_{\bot }^{a}}+\pi _{\Vert }^{ak}\overset{\rightharpoonup }{x}\times
\nabla A_{\bot k}^{a}
\end{equation*}%
\begin{equation}
+A_{\bot k}^{a}\varepsilon ^{lmn}x_{n}\overset{\rightharpoonup }{e_{l}}%
(D_{\Vert }^{k}\pi _{\Vert m}^{a}-D_{\Vert m}\pi _{\Vert }^{ak}),  \tag{4.11}
\end{equation}%
where $\overset{\rightharpoonup }{e_{l}}$ is the spatial unit vector and $%
D_{\Vert }^{k}\pi _{\Vert m}^{a}=\partial ^{k}\pi _{\Vert
m}^{a}+gf^{abc}A_{\Vert }^{bk}\pi _{\Vert m}^{c},$because there is
generalizing the Abelian Eq.(3.21) to non-Abelian gauge expression (3.24).
Adding Eqs.(3.17) and (3.18) to Eq.(4.3), we deduce Chen et al's useful
separation of the angular momentum \cite{6}
\begin{eqnarray*}
\overset{\rightharpoonup }{J_{3}} &=&\int d^{3}x\psi ^{\dagger }\frac{1}{2}%
\overset{\rightharpoonup }{\Sigma }\psi +\int d^{3}x\psi ^{\dagger }\overset{%
\rightharpoonup }{x}\times \frac{1}{i}\overset{\rightharpoonup }{D_{\Vert }}%
\psi \\
&&+\int d^{3}x\overset{\rightharpoonup }{\pi ^{a}}\times \overset{%
\rightharpoonup }{A_{\bot }^{a}}+\int d^{3}x\pi _{k}^{a}\overset{%
\rightharpoonup }{x}\times \nabla A_{\bot }^{ak}
\end{eqnarray*}%
\begin{equation}
=\overset{\rightharpoonup }{S^{q}}+\overset{\rightharpoonup }{L_{3}^{q}}+%
\overset{\rightharpoonup }{S_{3}^{g}}+\overset{\rightharpoonup }{L_{3}^{g}}.
\tag{4.12}
\end{equation}%
Therefore, in general field theory, we prove that Eq. (4.12) can strictly
and naturally be given out from the action (3.17) by Noether theorem.
Analogous to the derivation of Eq. (4.5), the gauge invariance of $\overset{%
\rightharpoonup }{L_{3}^{g}}$ demands the condition (4.7) (or Eqs. (4.6) and
(3.23)), which shows a key role in transforming Eq. (4.12) to the
conventional form \cite{6}
\begin{eqnarray*}
\overset{\rightharpoonup }{J_{4}} &=&\int d^{3}x\psi ^{\dagger }\frac{1}{2}%
\overset{\rightharpoonup }{\Sigma }\psi +\int d^{3}x\psi ^{\dagger }\overset{%
\rightharpoonup }{x}\times \frac{1}{i}\overset{\rightharpoonup }{\nabla }\psi
\\
&&+\int d^{3}x\overset{\rightharpoonup }{\pi ^{a}}\times \overset{%
\rightharpoonup }{A^{a}}+\int d^{3}x\pi _{k}^{a}\overset{\rightharpoonup }{x}%
\times \nabla A^{ak}
\end{eqnarray*}%
\begin{equation}
=\overset{\rightharpoonup }{S^{q}}+\overset{\rightharpoonup }{L_{4}^{q}}+%
\overset{\rightharpoonup }{S_{4}^{g}}+\overset{\rightharpoonup }{L_{4}^{g}},
\tag{4.13}
\end{equation}%
Consequently, under certain conditions, we can see that $\overset{%
\rightharpoonup }{J_{2}}$, $\overset{\rightharpoonup }{J_{3}}$ and $\overset{%
\rightharpoonup }{J_{4}}$ present the same value of the total angular
momenta. However, the last three terms of $\overset{\rightharpoonup }{J_{4}}$
are apparently all gauge dependent, thus $\overset{\rightharpoonup }{J_{4}}$
is not properly defined. In comparison, both $\overset{\rightharpoonup }{%
J_{2}}$ and $\overset{\rightharpoonup }{J_{3}}$ are more physical.
Furthermore, $\overset{\rightharpoonup }{J_{2}}$ is the simplest and the
most rational one for the sake of simplicity, because $\overset{%
\rightharpoonup }{J_{2}}$ eliminates the terms that don't contribute to the
total angular momentum for large $N_{c}$ gauge field theory, as Eq. (4.11)
shows.

5. Discussion

The nonlinear properties of interaction systems of the non-Abelian
quark-gluon fields produce considerable complex properties, and the
consistency and gauge invariance of the quark-gluon interaction system
theory naturally require to establish the six equations: (3.12), (3.13),
(3.23-3.25) and (4.9), which are directly uniformly generalized from their
Abelian expressions to their non-Abelian corresponding expressions according
to their gauge covariance and consistent properties, thus they can be
directly simplified as the Abelian corresponding expressions when their
structure constants $f^{abc}$ are taken as zero. In order to give very
obviously direct physics meanings of the six equations we rewrite them as
the direct compact expressions

\begin{equation}
\overrightarrow{D}_{\Vert }\cdot \overrightarrow{A}_{\bot }^{a}=0,  \tag{5.1}
\end{equation}%
\begin{equation}
(\overrightarrow{D}_{\Vert }\times \overrightarrow{D}_{\Vert })^{a}=0,
\tag{5.2}
\end{equation}%
\begin{equation}
\overrightarrow{D}_{\Vert }\cdot \overrightarrow{\pi }_{\bot }^{a}=0,
\tag{5.3}
\end{equation}%
\begin{equation}
\overrightarrow{D}_{\Vert }\times \overrightarrow{\pi }_{\Vert }^{a}=0,
\tag{5.4}
\end{equation}%
\begin{equation}
\overrightarrow{D}_{\Vert }\cdot \overrightarrow{\pi }_{\Vert }^{a}=\rho
^{a},  \tag{5.5}
\end{equation}%
\begin{equation}
\int d^{3}x\overrightarrow{\pi }_{\Vert }^{a}\cdot \overrightarrow{\pi }%
_{\bot }^{a}=0.  \tag{5.6}
\end{equation}

Given an arbitrary gauge field configuration $A_{\nu }(x)$, one can really
find an explicit solution $\overrightarrow{A}_{\bot }$ ($\overset{%
\rightharpoonup }{A_{\Vert }}=\overset{\rightharpoonup }{A}-\overrightarrow{A%
}_{\bot }$ is not independent), which satisfies all the six consistent
conditions (5.1-5.6), which gurrantee the consistency of the achieved theory
in this paper. Otherwise any new theory cannot return to the well-known
Abelian theory \cite{18}, which leads to no consistent theory.

Generalization from Abelian Eqs.(3.10) and (3.11) to non-Abelian Eq.(3.12)
and $\partial ^{j}A_{\Vert }^{ak}-\partial ^{k}A_{\Vert
}^{aj}+gf^{abc}A_{\Vert }^{bj}A_{\Vert }^{ck}=0$ and their important physics
meanings, respectively, are given by Wang et al \cite{14}. This paper also
gives the better and more direct expression (3.13) or (5.2) for non-Abelian
quark-gluon interaction system in general case than that in Ref.\cite{13}.
Consequently, the compact Eqs.(5.2-5.6) with direct physics meanings for a
general large $N_{c}$ QCD system in a general field theory are discovered in
this paper. The non-Abelian Eqs.(3.23) and (3.24), generalized from the
Abelian Eqs. (3.20) and (3.21), are the corresponding non-Abelian divergence
and curl generalizations, respectively (namely, the non-Abelian transverse
and parallel conditions); Eq. (5.5) is generalized from Eq.(3.22) and is the
new $SU(N_{c})$ Coulomb law; the orthogonal relation (5.6) is the
generalized expression for the Abelian $\int d^{3}x\overrightarrow{E}_{\Vert
}^{a}\cdot \overrightarrow{E}_{\bot }^{a}=0$. Furthermore, we show also that
Eq.(4.7) with a similar expression, imposed by Chen et al \cite{6}, which is
key to maintain $\overset{\rightharpoonup }{L_{3}^{g}}$ gauge invariant, is
just a direct result of Eqs.(3.23), (3.25) and (4.6). Under the six
consistent conditions, we naturally achieve the simplest gauge invariant
expression (4.3) of the total angular momentum of a general QCD system with
a general $N_{c}$---the general quark-gluon field interaction system by
Noether theorem in general field theory.

For all the above general investigations with a general $N_{c}$, we get a
general theory of general large $N_{c}$ QCD's inner structure and its gauge
invariant angular momentum and the corresponding new general Coulomb theorem
etc so that we give the new viewpoint on the resolution of the nucleon spin
crisis relative to the large $N_{c}$ in a general situation. Specially, this
theory is also suitable for investigating lattice QCD. For instance, one can
use this method to discuss the theory of the lattice QCD with large $N_{c}$
\cite{16}.

When the $N_{c}=3,$ it follows from Eq.(2.6) that

\begin{equation}
f^{ace}f^{bde}=\frac{2}{3}(\delta ^{ab}\delta ^{c}{}^{d}-\delta ^{ad}\delta
^{b}{}^{c})+d^{abe}d^{c}{}^{de}-d^{ade}d^{b}{}^{ce}  \tag{5.7}
\end{equation}%
i.e., the current QCD \cite{16} is considered. Thus substituting Eq.(5.7)
into all its relevant expressions in all the above investigations, then we
get a general theory of QCD's inner structure and its gauge invariant
angular momentum as well as the corresponding new Coulomb theorem etc so
that we give out the new viewpoint on the resolution of the nucleon spin
crisis in a general field theory. The gauge invariance and angular momenta
commutation relation are not satisfied simultaneously in the frame of
general field theory of QCD in the past, however, this paper, for the first
time, makes these satisfied in the same time in a general field theory,
without artificial choice.

When the $N_{c}=2,$ Eq.(2.6) is reduced as

\begin{equation}
f^{ace}f^{bde}=\delta ^{ab}\delta ^{c}{}^{d}-\delta ^{ad}\delta ^{b}{}^{c},
\tag{5.8}
\end{equation}%
because $d^{abe}$ can only be equivalently taken as $d^{123}$ here, then
replacing the quark-gluon interaction with the fermi-gauge interaction, the
above theory is simplified as the Ref.\cite{13}'s investigations and which
are the exact same as Ref.\cite{13} results, i.e., the current
hadron-dynamics gauge field theory about strong interaction with different
charged electricity systems \cite{13} is achieved.

When the $N_{c}=1,$people usually take U(1) gauge symmetry, i.e., the
Abelian gauge field is considered, its structure constant is zero, thus
replacing the quark-gluon interaction with the fermi-gauge interaction, the
above theory is simplified as the Ref.\cite{12}'s investigations and which
are the exact same as Ref.\cite{12}'s results.

6. Summary and conclusion

Consequently, the above theory not only gives the general intrinsic theory
of general large $N_{c}$ QCD, SU(3) QCD, SU(2) hadron-dynamics and U(1) QED
gauge field theories in general field theory, but also and specially
meaningfully makes essential steps towards solving nucleon spin crisis in
the general field theory, i.e., presents progress towards solving the
nucleon spin crisis. The discoveries of the inner-structure similarity among
the above theories, their gauge invariant angular momenta and the
corresponding new Coulomb theorem etc of the non-Abelian fermi-gauge field
interaction systems with a general large $N_{c}$ vividly uncover the
consistency and gauge invariance of the gauge field theory and will further
deepen our understanding on some basic questions in physics.

Because of the universality of Eqs. (5.1)-(5.6), they can be applied to
treat the inner structure and other aspects of any fermi-gauge interaction
system, including SU(N) gauge field theory and different gauge theories for
the fundamental interactions of the universe. Their aspects like
Euler-Lagrange equations, corresponding conservation quantities and relevant
dynamics etc. need to be revised and improved by using the new general
theory of the decomposed non-Abelian gauge fields. Due to the length limit
of the paper, lots of applications of this paper's theory will be written in
our following works.

This paper not only presents the simplest expression (4.3) but also gives
the current expression (4.12), and discovers their relation between
Eqs.(4.3) and (4.12).

This paper decomposes the gluon fields into transverse and longitudinal
parts in general field theory, which are then used to obtain a general
expression for the total angular momentum $J$ of a general quark-gluon
system in general field theory. This expression divides $J$ into quark spins
and gluon spins and their orbital angular momentum parts in general field
theory, with gauge invariant expressions for each of the contributions.

This paper is one in a long history of attempting to decompose the nucelon's
spin into various pieces which together must sum to one half. Such an effort
is really of general interest because if the individual parts are either
measurable experimentally or directly computable in the physics. This paper
just gives the theory as how the individual parts of the spin in the
decomposition can be obtained theoretically and experimentally because all
the achieved quantities are exactly calculated and gauge invariant, which
then results in that they are measurable, because the quantities without
gauge invariance cannot be exactly measured \cite{6}.

This paper's generally decomposing gauge potential theory also presents a
new technique, or methodology, it should play a votal role in future physics
research, and make apparent its immediate consequences for physicists,
because the generally decomposing gauge potential theory in general field
theory can be directly and effectively applied to arbitrary SU(N), SO(N) etc
gauge interaction systems with different Fermi and the other matter fields.

Therefore, this paper makes a breakthrough progress in overcoming the key
huge difficulty of nucleon spin crisis, i.e. succeeding in figuring out a
natural way to construct, for the first time, the consistent angular
momentum operators with both their commutation relation and gauge invariance
satisfied in general field theory, and further a new door is eventually
opened to investigate lots of strong interacting systems with different
symmetric properties, which is very important in understanding the nature
and thus popularly interesting, and their theories are consistent. These
interesting works here will be extensively applied and largely cited, and
further be written into relevant books because of their generalities and
being able to be extensively applied.

Specially, we want to stress that the achieved theory in this paper can be
utilized to calculate the general QCD strong interactions and give the
precise predictions, further the achieved predictions in the calculations
can be exactly measured by current particle physics experiments due to their
gauge invariant properties etc, because any physical quantity ( relative to
gauge transformations ) without gauge invariant property cannot be exactly
measured \cite{6}.

ACKNOWLEDGMENT: This work is partly supported by NSF through a CAREER Award
PHY-0952630 and by DOE through Grant DE-SC0007884 in USA and the National
Natural Science Foundation of China (Grants No. 11275017 and No. 11173028).

\bibliographystyle{plain}
\bibliography{paperQCD1}

\begin{thebibliography}{99}
\bibitem{1} C. N. Yang and R. Mills, Phys. Rev. 96 (1): 191.

\bibitem{2} Thomas D. Cohen, Brian Mc Peak, Bendeguz Offertaler, Phys. Rev.
C 92, 015204 (2015).

\bibitem{3} E. Ageev et al. (COMPASS Collaboration), Physics Letters B 612,
154 (2005).

\bibitem{4} V. Alexakhin et al. (COMPASS Collaboration), Physics Letters B
647, 8 (2007).

\bibitem{5} A. Airapetian et al. (HERMES Collaboration), Phys. Rev. D 75,
012007 (2007).

\bibitem{6} X. S. Chen, X. F. Lu, W. M. Sun, F. Wang, and T. Goldman, Phys.
Rev. Lett. 100, 232002 (2008).

\bibitem{7} X. Ji, Phys. Rev. Lett. 78, 610 (1997).

\bibitem{8} M. Wakamatsu, Phys. Rev. D81, 114010 (2010), arXiv: 1004.0268
[hep-ph].

\bibitem{9} X.~Ji, Phys.\ Rev.\ Lett.\ \textbf{104}, 039101 (2010); X.~Ji,
Phys.\ Rev.\ Lett.\ \textbf{106}, 259101 (2011) [arXiv:0910.5022 [hep-ph]]

\bibitem{10} X.~Ji, J.~H.~Zhang and Y.~Zhao, Phys.\ Rev.\ Lett.\ \textbf{111}%
, 112002 (2013) [arXiv:1304.6708 [hep-ph]].

\bibitem{11} Y.~Hatta, X.~Ji and Y.~Zhao, Phys.\ Rev.\ D \textbf{89}, no. 8,
085030 (2014) [arXiv:1310.4263 [hep-ph]]; X.~Ji, J.~H.~Zhang and Y.~Zhao,
Phys.\ Lett.\ B \textbf{743}, 180 (2015) [arXiv:1409.6329 [hep-ph]].

\bibitem{12} B. H. Zhou and Y. C. Huang, Phys. Rev. D84, 047701 (2011);
Phys. Rev. A84, 032505 (2011) .

\bibitem{13} C. Huang, Y. C. Huang and B. H. Zhou, Phys. Rev. D 92, 056003
(2015).

\bibitem{14} F. Wang, X. S. Chen, X. F. Lu, W. M. Sun, and T. Goldman,
(2009), arXiv:0909.0798 [hep-ph].

\bibitem{15} S. Lerma H., B. Errea, J. Dukelsky et al, Phys. Rev. Lett., 99,
032501 (2007); Sz. Borsanyi, S. D\"{u}rr, Z. Fodor et al, Phys. Rev. Lett.,
111, 252001 (2013).

\bibitem{16} Review of Particle Physics, Phys. Rev. D 86 (2012) 010001.

\bibitem{17} W. Greiner and J. Reinhardt, Field Quantization (Beijing World
Publishing Corp., 2003).

\bibitem{18} M. E. Peskin and D. V. Schroeder, An Introduction to Quantum
Field Theory (Beijing World Publishing Corp., 2006).
\end{thebibliography}

\end{document}